\documentclass{PoS}

\def\simge{\mathrel{%
       \rlap{\raise 0.511ex \hbox{$>$}}{\lower 0.511ex \hbox{$\sim$}}}}
\def\simle{\mathrel{
       \rlap{\raise 0.511ex \hbox{$<$}}{\lower 0.511ex \hbox{$\sim$}}}}

\title{Study of $2+1$ flavor finite-temperature QCD using improved Wilson quarks at the physical point with the gradient flow%
\\ \vspace*{-54mm}\hspace{7cm} \small{\textrm{UTHEP-736, J-PARC-TH-0171, KYUSHU-HET-200}} \vspace*{52mm}
}

\ShortTitle{$2+1$ flavor finite temperature QCD with gradient flow}

\author{\speaker{Kazuyuki Kanaya}\\ 
        Tomonaga Center for the History of the Universe, 
        Univ. of Tsukuba, Tsukuba 305-8571, Japan\\
        E-mail: \email{kanaya@ccs.tsukuba.ac.jp}}

\author{Atsushi Baba, Asobu Suzuki\\
        Graduate School of Pure and Applied Sciences, Univ.\ of Tsukuba, Tsukuba 305-8571, Japan
        }

\author{Shinji Ejiri\\
        Department of Physics, Niigata Univ., Niigata 950-2181, Japan
        }

\author{Masakiyo Kitazawa\\
        Department of Physics, Osaka Univ., Osaka 560-0043, Japan\\
	J-PARC Branch, KEK Theory Center, Institute of Particle and Nuclear Studies, KEK, 203-1, Shirakata, Tokai, Ibaraki 319-1106, Japan
        }

\author{Hiroshi Suzuki\\
        Department of Physics, Kyushu Univ., 744 Motooka, Nishi-ku, Fukuoka 819-0395, Japan
        }

\author{Yusuke Taniguchi\\
        Center for Computational Sciences, Univ.\ of Tsukuba, Tsukuba 305-8577, Japan
        }

\author{Takashi Umeda\\
        Graduate School of Education, Hiroshima Univ., Hiroshima 739-8524, Japan
        }

\abstract{We study thermodynamic properties of $2+1$ flavor QCD applying the \textbf{S}mall \textbf{F}low-\textbf{\textit{t}}ime e\textbf{X}pansion (\textbf{SF\textit{t}X}) method based on the gradient flow. 
The method provides us with a general way to compute correctly renormalized observables irrespective of explicit violation of symmetries due to the regularization, such as the Poincar\'e and chiral symmetries on the lattice.
We report on the status of our on-going project to compute the energy-momentum tensor and the chiral condensate at the physical point with improved Wilson quarks,
extending our previous study with slightly heavy $u$ and $d$ quarks.  
We also report on our test of two-loop matching coefficients recently calculated by Harlander \textit{et al.},
revisiting the case of QCD with slightly heavy $u$ and $d$ quarks. 
Our results suggest that the SF\textit{t}X method is powerful in extracting physical observables on the lattice.
}

\FullConference{37th International Symposium on Lattice Field Theory - Lattice2019\\
		16-22 June 2019\\
		Wuhan, China}

\begin{document}

\section{Introduction}

The gradient flow (GF) opened us a variety of new methods to significantly simplify the calculation of physical observables on the lattice~\cite{Narayanan:2006rf,Luscher:2010iy}.
Among them, we are applying the \textbf{Small Flow-\textit{t}ime eXpansion} (\textbf{SF\textit{t}X}) \textbf{method}~\cite{Suzuki:2013gza} based on the GF, 
which is a general method  to correctly calculate any renormalized observables on the lattice provided that the correct continuum limit is guaranteed as in the case of QCD with Wilson-type quarks. 
The method was first applied to calculate the energy-momentum tensor (EMT), which has been not easy to evaluate on the lattice due to explicit violation of the Poincar\'e invariance at finite lattice spacings \cite{Makino:2014taa}. 
The method was shown to be powerful by a test in quenched QCD~\cite{FlowQCD1}.
We note that the SF\textit{t}X method is applicable also to observables related to the chiral symmetry~\cite{Hieda:2016lly}. 
We thus apply the method to QCD with Wilson-type quarks to cope with the problems due to explicit violation of the chiral symmetry by the Wilson term.
To reduce the finite lattice spacing effects, we adopt the renormalization-group improved Iwasaki gauge action and the non-pertuebatively $O(a)$-improved Wilson quark action. 

We have first studied the case of $2+1$ flavor QCD with slightly heavy $u$ and $d$ quarks ($m_\pi/m_\rho\simeq0.63$) while the $s$ quark mass is approximately its physical value ($m_{\eta_{ss}}/m_\phi\simeq0.74$)~\cite{Taniguchi:2016ofw,Taniguchi:2016tjc}.
On a relatively fine lattice with the lattice spacing $a=0.07$ fm \cite{Ishikawa:2007nn}, finite-temperature configurations in the range $T\simeq174$--697 MeV ($N_t=16$--4, where $N_t$ is the lattice size in the temporal direction) are generated based on the fixed-scale approach \cite{Umeda:2008bd}.
At $T\simle280\,\mathrm{MeV}$ ($N_t \simge 10$), we found that the equation of state (EoS) extracted from diagonal components of EMT by the SF\textit{t}X method~\cite{Taniguchi:2016ofw} is consistent with our previous estimation using the conventional $T$-integration method on the same configurations~\cite{Umeda:2012er}.
At the same time, the two estimates of EoS deviate at $T\simge350\,\mathrm{MeV}$, suggesting contamination of $a$-independent lattice artifacts of $O\!\left((aT)^2\right)=O\!\left(1/N_t^2\right)$ at $N_t \simle 8$.
We also studied the chiral condensates and its susceptibilities. 
We found that, in spite of the explicit chiral violation by Wilson-type quarks, the chiral condensates and the chiral susceptibilities show signals of transition/crossover just at the pseudo-critical temperature suggested from other observables~\cite{Taniguchi:2016ofw}. 
We have further studied topological properties of QCD on these lattices adopting the SF\textit{t}X method~\cite{Taniguchi:2016tjc}.
We found that the topological susceptibility estimated with the gluonic and fermionic definitions agree well with each other at $T\simle 280$ MeV already on our finite lattices.
This is in clear contrast to the conventional lattice estimations of them: 
For example, a study with improved staggered quarks reports more than hundred times larger gluonic susceptibility than fermionic one at similar lattice spacings~\cite{Petreczky:2016vrs}.
These suggest that the SF\textit{t}X method is powerful in extracting physical observables from lattice simulations.

In Sec.~\ref{sec:physpt}, we report on the status of our extension of this study to  $2+1$ flavor QCD with physical $u$, $d$, and $s$ quarks. 
In Sec.~\ref{sec:twoloop}, we then report on our test of two-loop matching coefficients of the SF\textit{t}X method, recently calculated in Ref.~\cite{HKL}, revisiting the case with slightly heavy $u$ and $d$ quarks.
A summary and prospects are given in Sec.~\ref{sec:summary}. 

\section{QCD at the physical point}
\label{sec:physpt}

We are extending the study of $2+1$ flavor QCD in Ref.~\cite{Taniguchi:2016ofw} to the case with physical $u$, $d$, and $s$ quarks~\cite{physicalpoint}, using the same combination of gauge and quark actions.
The quark masses are fine-tuned to the physical point on a lattice with lattice spacing of 0.09 fm~\cite{Aoki:2009ix}. 
Finite temperature simulations are done in the temperature range $T\simeq122$--548 MeV ($N_t=18$--4) with the spatial box size of $32^3$, while, according to our experience of Ref.~\cite{Taniguchi:2016ofw}, EMT data at $T\simge274\,\mathrm{MeV}$ will be contaminated with $a$-independent $O\!\left((aT)^2 = 1/N_t^2\right)$ lattice artifacts at $N_t \simle 8$.

Thanks to the asymptotic freedom around the small-$t$ limit, we can apply the perturbation theory to calculate the small-$t$ expansion of operators in the GF-scheme (``flowed operators'') in terms of the running coupling and masses in a conventional renormalization scheme, say, the $\overline\textrm{MS}$-scheme, at renormalization scale $\mu$.
A conventional choice of $\mu$ is $\mu=\mu_d(t) \equiv 1/\sqrt{8t}$ which is a natural scale of flowed observables because the GF affects as a smearing of fields over a physical range of $\sqrt{8t}$.
By inverting the relation, we can expand renormalized observables in terms of flowed operators at small~$t$.
Contamination of higher order terms in $t$ can be removed by a $t\to0$ extrapolation.
Following Ref.~\cite{Taniguchi:2016ofw}, we combine five lattice operators using the one-loop matching coefficients of~\cite{Makino:2014taa} and calculate the correctly renormalized EMT by performing a $t\to0$ extrapolation of it using data within a linear window.

In our previous test at the physical point using the $\mu_d$-scale, we found that several linear windows became narrower than those in the case of slightly heavy $u$ and $d$ quarks~\cite{Taniguchi:2016ofw}, that sometimes makes the linear $t\to0$ extrapolation ambiguous~\cite{physicalpoint}.
A reason behind this is that the lattice is slightly coarser. 
A larger lattice spacing $a$ limits the meaningful range of $t/a^2$ narrower from above due to the IR explosion of the running coupling at large~$t$.

We note that this limitation can be relaxed by choosing the renormalization scale $\mu$ in the matching coefficients appropriately: 
Harlander \textit{et al.} argued that the scale $\mu_0(t) \equiv 1/\sqrt{2e^{\gamma_E}t}$, which simplifies perturbative expressions for matching coefficients, also keeps the two-loop contributions small~\cite{HKL}. Here, $\gamma_E$ is the Euler-Mascheroni constant.
As far as the perturbative expansions for the matching coefficients are under control, the final physical observables should be insensitive to the choice of~$\mu$.
In practice of lattice analyses, however, because $\mu_0(t) \simeq 1.5 \times \mu_d(t)$, the $\mu_0$-scale improves the quality of perturbative expansion and extends the perturbative range towards larger~$t$.

\begin{figure}[tb]
 \centering
  \includegraphics[width=5cm,trim= 0 1mm 7mm 0]{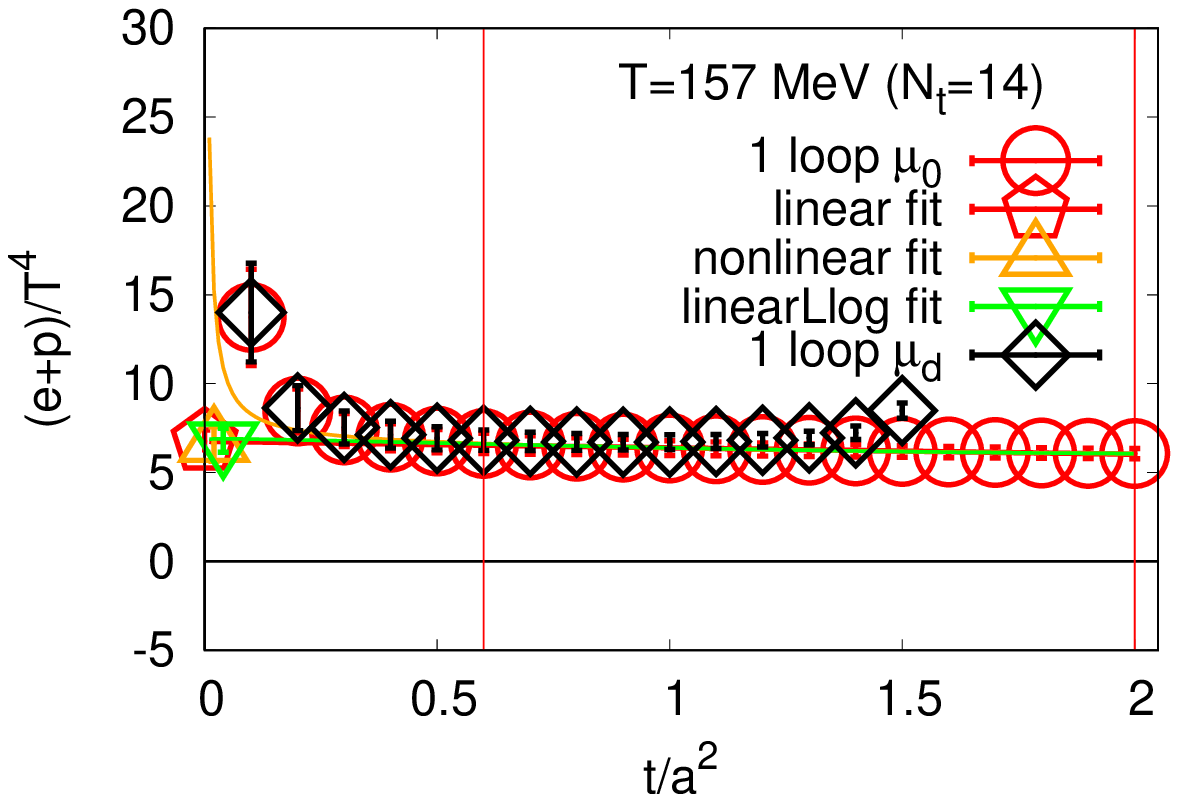}
  \includegraphics[width=4.8cm,trim=0 0 7mm 6mm]{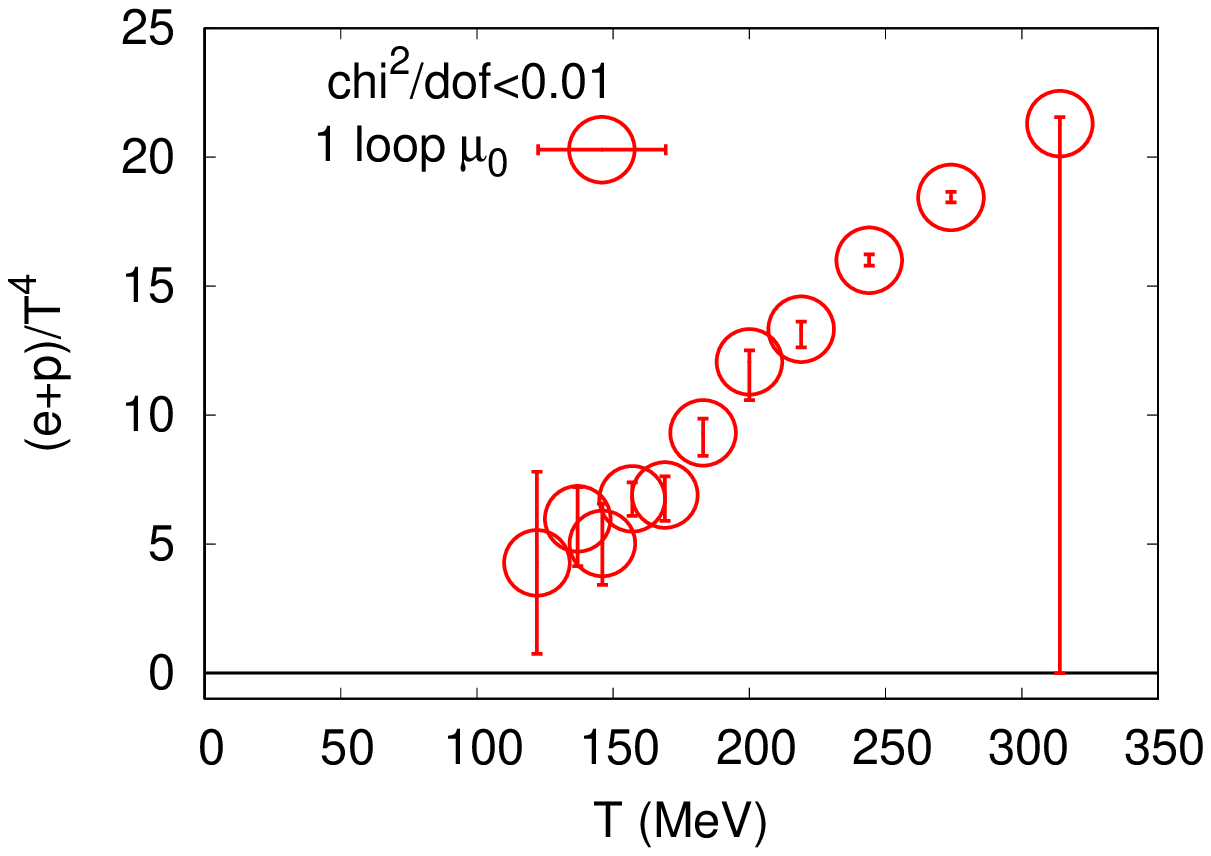}
  \includegraphics[width=4.8cm,trim=0 0 7mm 6mm]{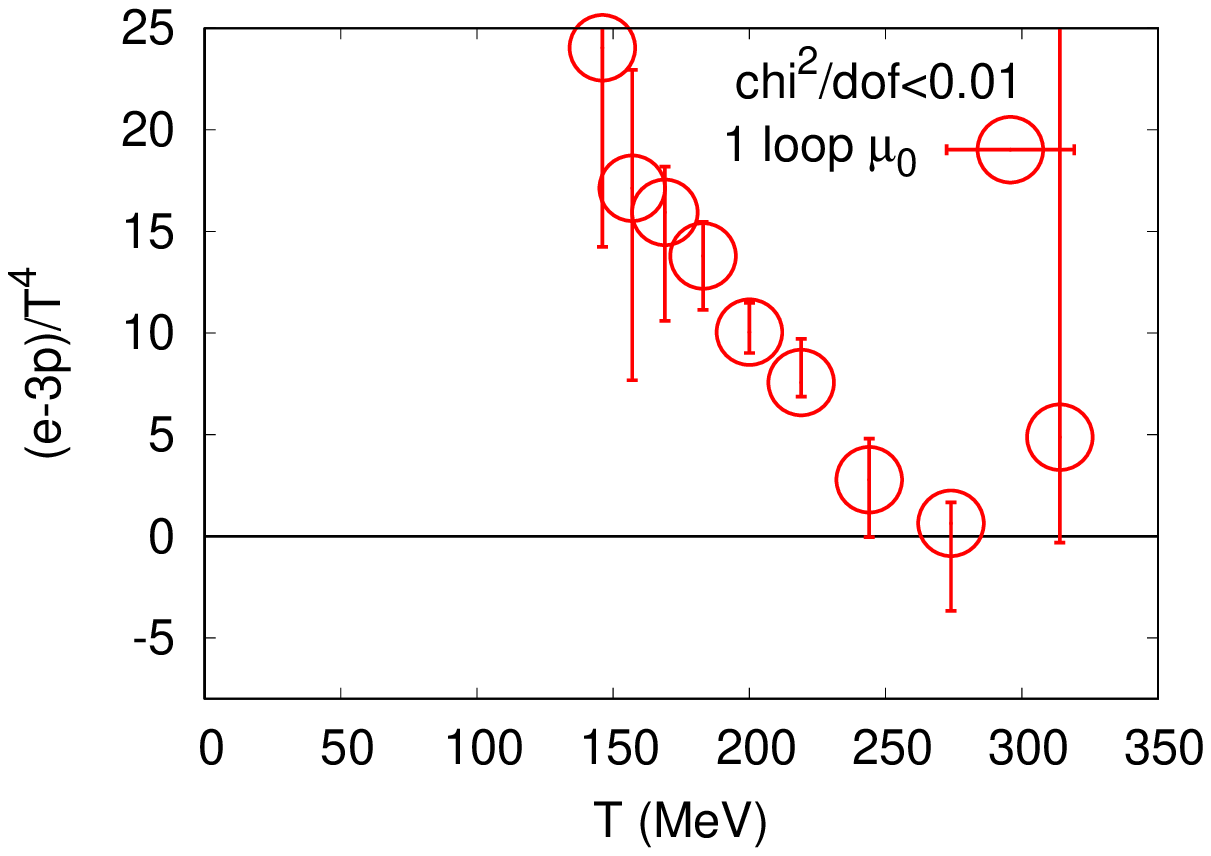}
  \vspace*{-2mm}
  \caption{
  EoS in QCD at the physical point.
  \textbf{Left}: Entropy density at $T\simeq 157$ MeV as a function of the flow-time. 
  Red circles and black diamonds are the results of $\mu_0$- and $\mu_d$-scales, respectively.
  Also shown are the results of various $t\to0$ extrapolations using the $\mu_0$-scale data within the linear window shown by two thin vertical lines. 
  Two thin vertical lines show the linear window for the $\mu_0$-scale.
  Errors are statistical only.
  \textbf{Middle}:  Entropy density in the $t\to0$ limit using the $\mu_0$-scale as a function of temperature. Errors include systematic errors from the $t\to0$ extrapolation.
  Data at $T\simge274\,\mathrm{MeV}$ are contaminated with $O\!\left((aT)^2=1/N_t^2\right)$ lattice artifacts at~$N_t \simle 8$.
  \textbf{Right}:  The same as the middle panel but for the trace anomaly.
}
\label{fig:physpt1}
\end{figure}

\begin{figure}[tb]
 \centering
  \includegraphics[width=4.9cm,trim=12mm 1mm 6mm 0,clip]{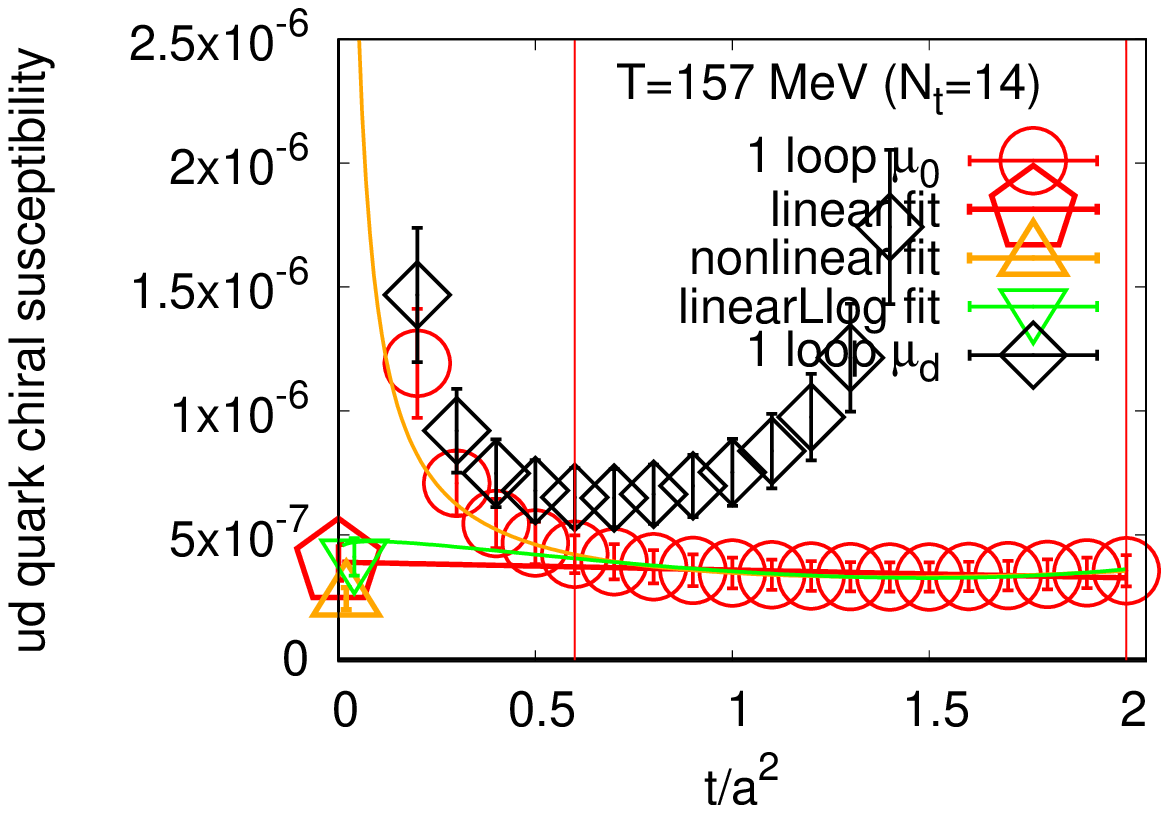}
  \includegraphics[width=4.7cm,trim=12mm 0 4mm 4mm,clip]{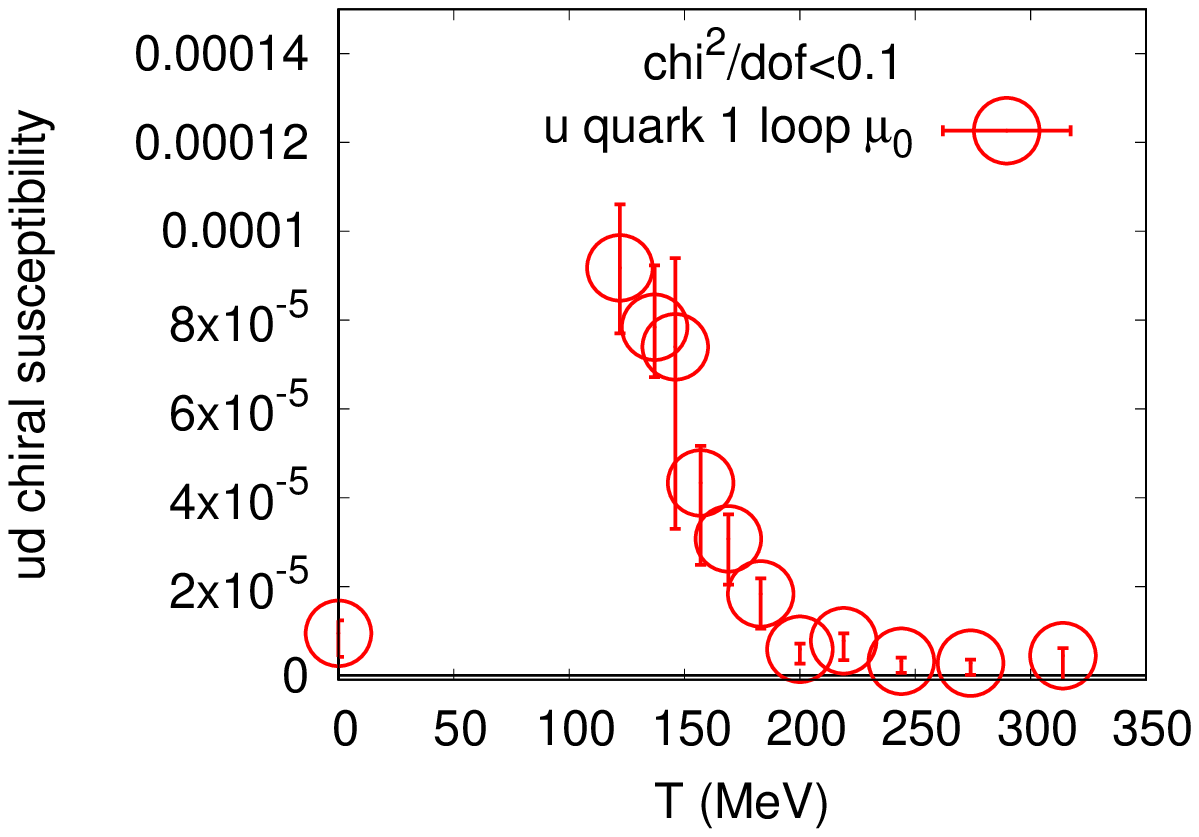}
  \includegraphics[width=4.7cm,trim=12mm 0 4mm 4mm,clip]{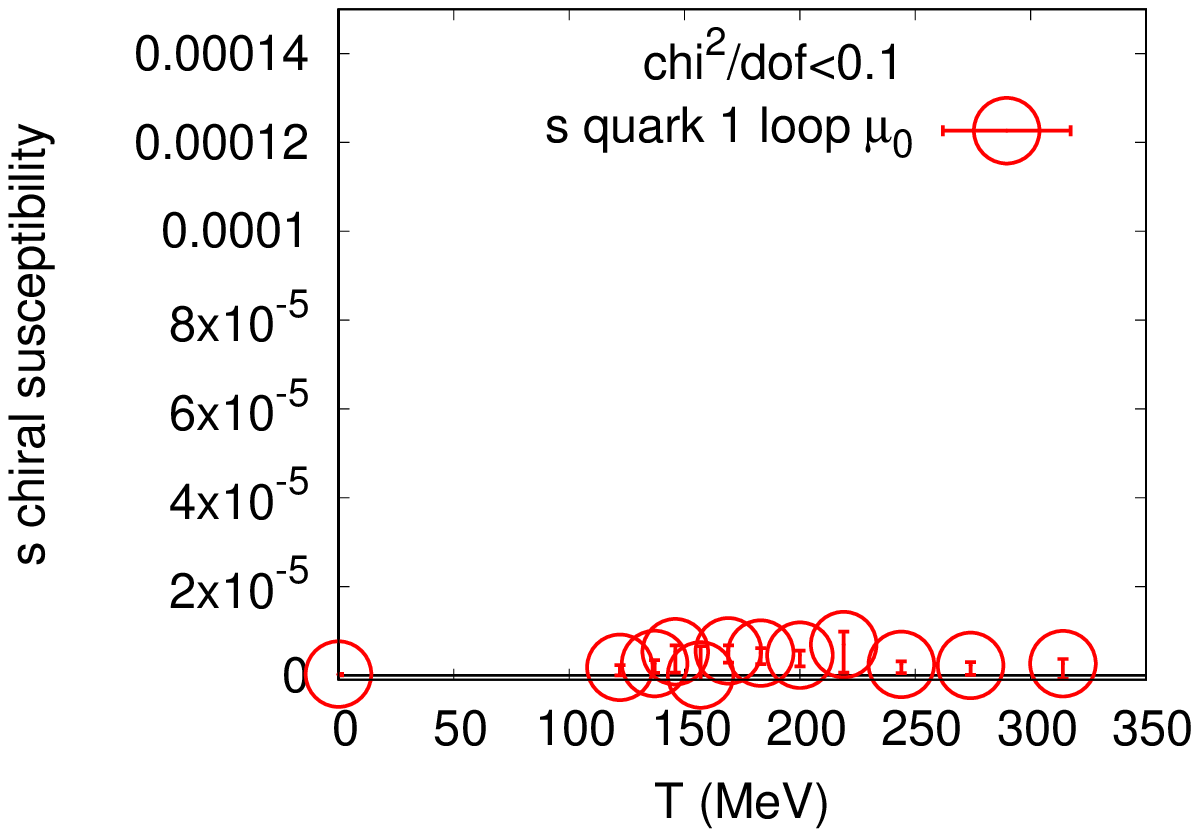}
  \vspace*{-2mm}
  \caption{
  Disconnected chiral susceptibilities in the $\overline\textrm{MS}$ scheme at 2 GeV in 2+1 flavor QCD at the physical point.
  \textbf{Left}: Chiral susceptibility of $u$ or $d$ quark. Results at $T\simeq 157$ MeV as a function of the flow-time. The vertical axis is in lattice unit. Errors are statistical only.
  \textbf{Middle}: Chiral susceptibility of $u$ or $d$ quark in the $t\to0$ limit with the $\mu_0$-scale. The vertical axis is in (GeV)$^6$. Errors include systematic errors from the $t\to0$ extrapolation.
  \textbf{Right}:  The same as the middle panel but for the $s$ quark.
}
\label{fig:physpt2}
\end{figure}

In the left panel of Fig.~\ref{fig:physpt1}, we show the entropy density as function of dimension-less flow time $t/a^2$. 
Red circles and black diamonds are the results of $\mu_0$- and $\mu_d$-scales, respectively.
With the conventional $\mu_d$-scale, $t/a^2$ is limited to be smaller than about 1.5 where the running coupling exceeds $O(1)$.
The range is extended up to above 3 by the $\mu_0$-scale.
With better control of the perturbative expansions by $\mu_0$, we now see clearer and wider linear windows so that we can carry out safe $t\to0$ extrapolations, similar to the case of slightly heavy $u$ and $d$ quarks studied in Ref.~\cite{Taniguchi:2016ofw}.
Our preliminary results for the EoS with the $\mu_0$-scale are shown in the middle and right panels of~Fig.~\ref{fig:physpt1}.
The simulations at $T\simeq122$--146 MeV are on-going yet.

In Fig.~\ref{fig:physpt2}, corresponding results for the disconnected chiral susceptibilities are shown. 
As shown in the left panel of Fig.~\ref{fig:physpt2}, an appropriate choice of $\mu$ turned out to be essential to get a clear linear window for this operator at these temperatures. 
Our preliminary results for the disconnected chiral susceptibilities in the $t\to0$ limit are shown in the middle and right panels for $u$ (or $d$) and $s$ quarks, respectively.
Though the statistics at $T\simeq122$--146 MeV are low yet, our current data suggests that the temperatures in $T\simeq 122$--146 MeV are in the critical region. 

\section{Test of two-loop matching coefficients}
\label{sec:twoloop}

We now turn to the topics of two-loop matching coefficients.
Harlander, Kluth and Lange have recently calculated the matching coefficients for EMT up to the two-loop order in~\cite{HKL}.
Removing more known small-$t$ behaviors, we may expect a milder $t$-dependence in the $t\rightarrow0$ extrapolations.
The two-loop coefficients were first tested in quenched QCD in Ref.~\cite{Iritani2019}.
They found that the results of EoS with one- and two-loop coefficients are well consistent with each other.
They also found that the two-loop coefficients lead to a milder $t$-dependence such that systematic errors from the $t\to0$ extrapolation are reduced.

We extend the test to full QCD.
A point to be noted in the full QCD coefficients is that, unlike the one-loop coefficients of Ref.~\cite{Makino:2014taa}, in the calculation of two-loop coefficients in \cite{HKL}, the equation of motion (EoM) in the continuum,
\begin{equation}
	\bar\psi(x)\left(\frac{1}{2}\stackrel{\leftrightarrow}{\not\!\!D} + m_0 \right) \psi(x) = 0,
\end{equation}
is used for quark operators, with which we can reduce the number of independent operators and coefficients for EMT.
This should cause no effects after taking the continuum limit when the EMT operators are isolated.
On finite lattices, however, EoM will get $O(a)$ lattice corrections that may induce additional systematic errors at $a\ne0$. 

\begin{figure}[tb]
 \centering
  \includegraphics[width=4.8cm,trim=0 0 9mm 8mm]{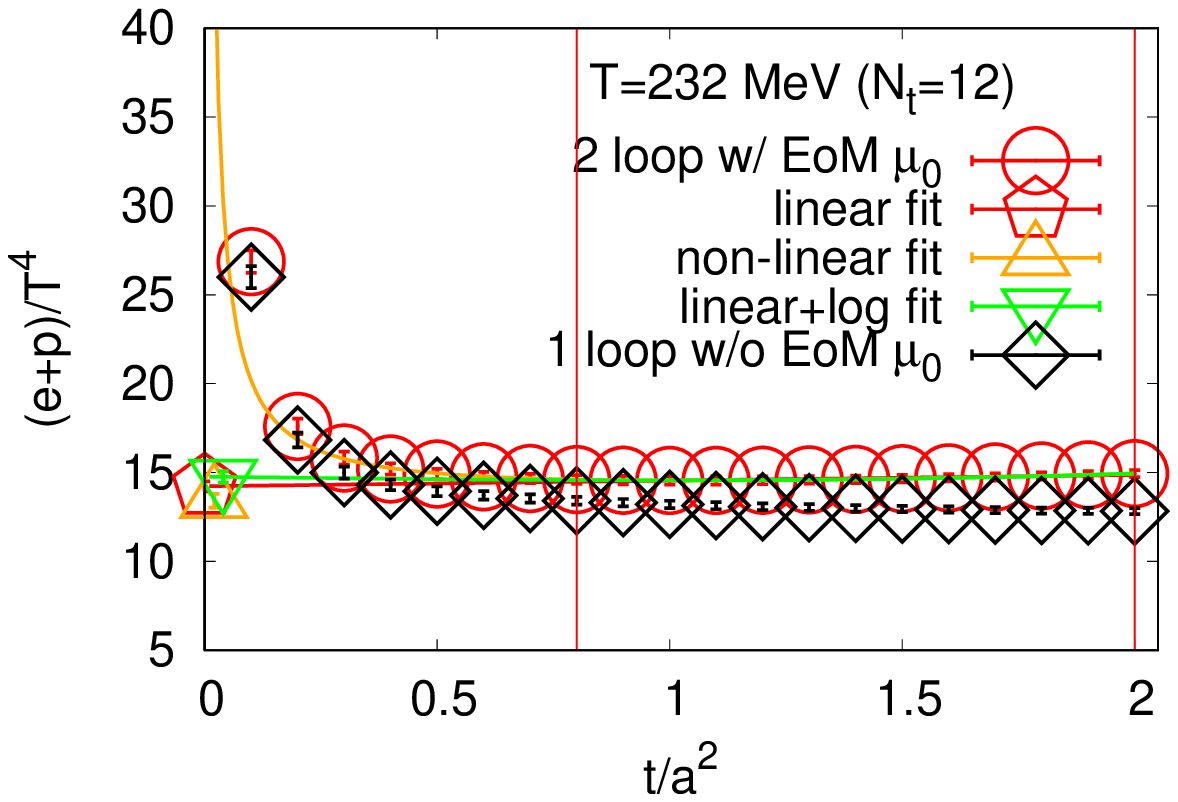}
  \includegraphics[width=5cm,trim=0 0 7mm 5mm]{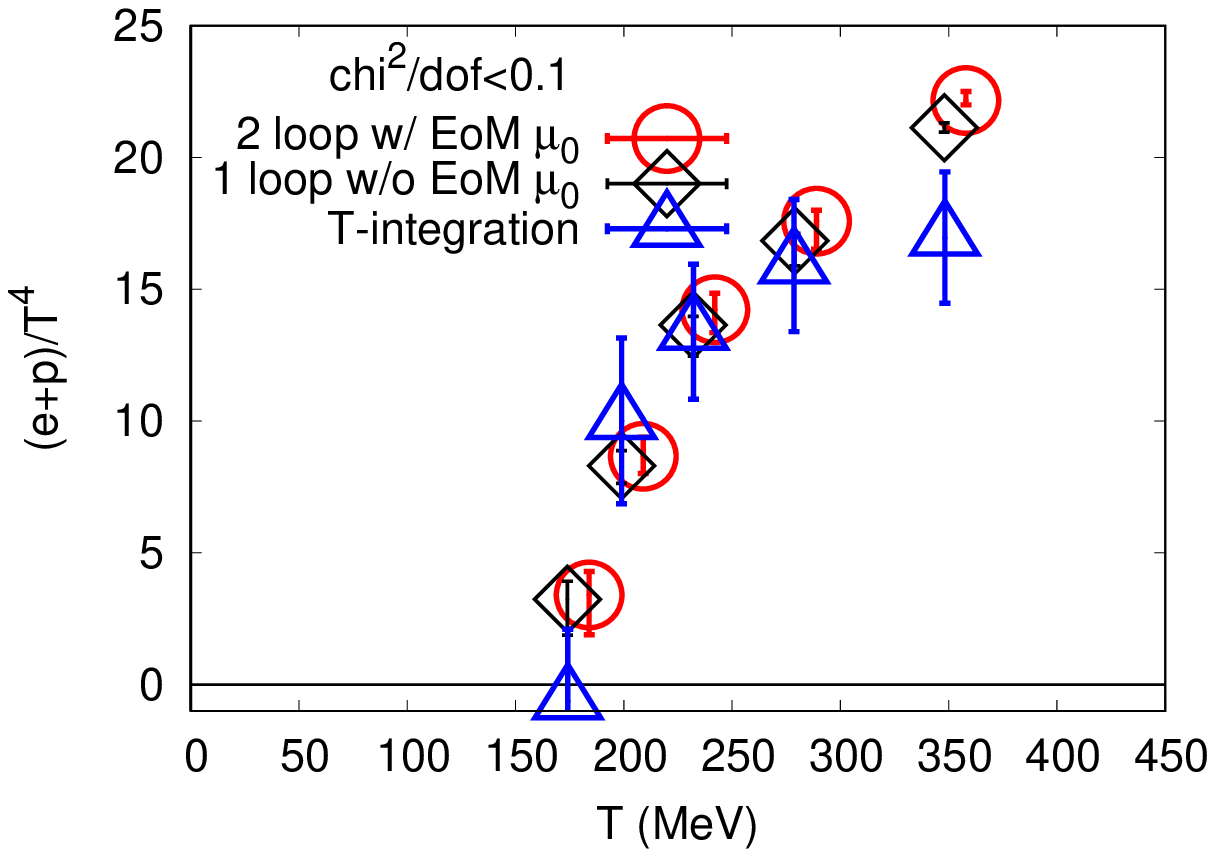}
  \includegraphics[width=5cm,trim=0 0 6mm 5mm]{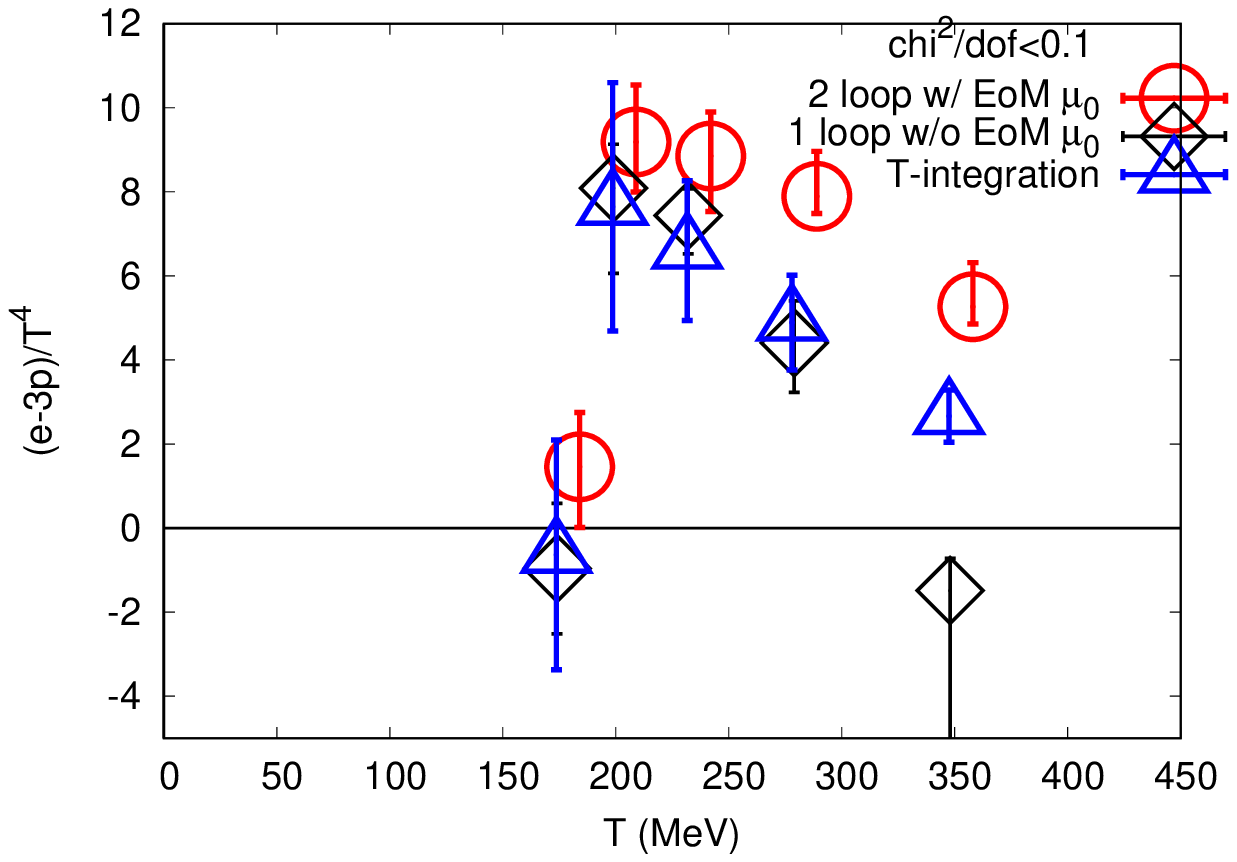}
  \vspace*{-2mm}
  \caption{
  EoS in QCD with slightly heavy $u$ and $d$ quarks using the $\mu_0$-scale. 
  Comparison of results using the one-loop matching coefficients of Ref.~\cite{Makino:2014taa} and those using the two-loop coefficients by Harlander \textit{et al.}~\cite{HKL}. EoM is used in the latter.
  Blue triangles are the results of the conventional $T$-integration method~\cite{Umeda:2012er}.
  \textbf{Left}: Entropy density at $T\simeq 232$ MeV as a function of the flow time. Errors are statistical only.
  \textbf{Middle}: Entropy density in the $t\to0$ limit. Errors include systematic errors from the $t\to0$ extrapolation.
  Data at $T\simge350\,\mathrm{MeV}$ are contaminated with $O\!\left((aT)^2=1/N_t^2\right)$ lattice artifacts at~$N_t \simle 8$.
  \textbf{Right}: The same as the middle panel but for the trace anomaly.
}
\label{fig:heavy}
\end{figure}

To test the effects of two-loop coefficients in full QCD, we revisit the case of slightly heavy $u$ and $d$ quarks.
Following the experience of Sec.~\ref{sec:physpt}, we adopt $\mu_0$ as the renormalization scale, 
though the improvement with the $\mu_0$-scale turned out to be not so drastic on this fine lattice.
The new results using the $\mu_0$-scale agree completely with our previous results using the $\mu_d$-scale \cite{Taniguchi:2016ofw}.

In the left and middle panels of Fig.~\ref{fig:heavy}, we compare the entropy density using the one-loop matching coefficients of~\cite{Makino:2014taa} (black diamonds) and that using the two-loop coefficients by Harlander \textit{et al.}~\cite{HKL} (red circles). 
Note that, because the EoM affects only the trace part of the EMT, the EoM has no effects in the entropy density which is a trace-less combination of the EMT.
In the left panel, we show the entropy density at $T\simeq 232$ MeV as a function of the flow time. 
We see small shifts at finite $t$ due to the use of the two-loop coefficients, but the difference vanishes in the $t\to0$ limit within the statistical errors.
Final results for the entropy density at all temperatures are shown in the middle panel of Fig.~\ref{fig:heavy}.
The errors contain systematic errors due to the $t\to0$ extrapolation as estimated in Ref.~\cite{Taniguchi:2016ofw} --- for linear+log fits in the two-loop case, the $\log^{-2} (\sqrt{8t}/a)$ term in~Eq.~(45) of~\cite{Taniguchi:2016ofw} is replaced by a~$\log^{-3} (\sqrt{8t}/a)$ term corresponding to the leading effects of~$O(g^6)$ higher order ambiguities.
We find that one- and two-loop results agree well with each other.

In the right panel of Fig.~\ref{fig:heavy}, we show the corresponding results for the trace anomaly, which is just the trace part of the EMT and thus will be sensitively affected by the EoM on finite lattices.
In this comparison, the EoM was used in the two-loop coefficients of Harlander \textit{et al.}, but was not used in the calculation of one-loop coefficients of Ref.~\cite{Makino:2014taa}.
We find that, though the two analyses are consistent within errors at low temperatures, 
they show visible discrepancy at $T\simge279$ MeV ($N_t\simle10$).

\begin{figure}[tb]
 \centering
  \includegraphics[width=4.9cm,trim=0 0 8mm 3mm,clip]{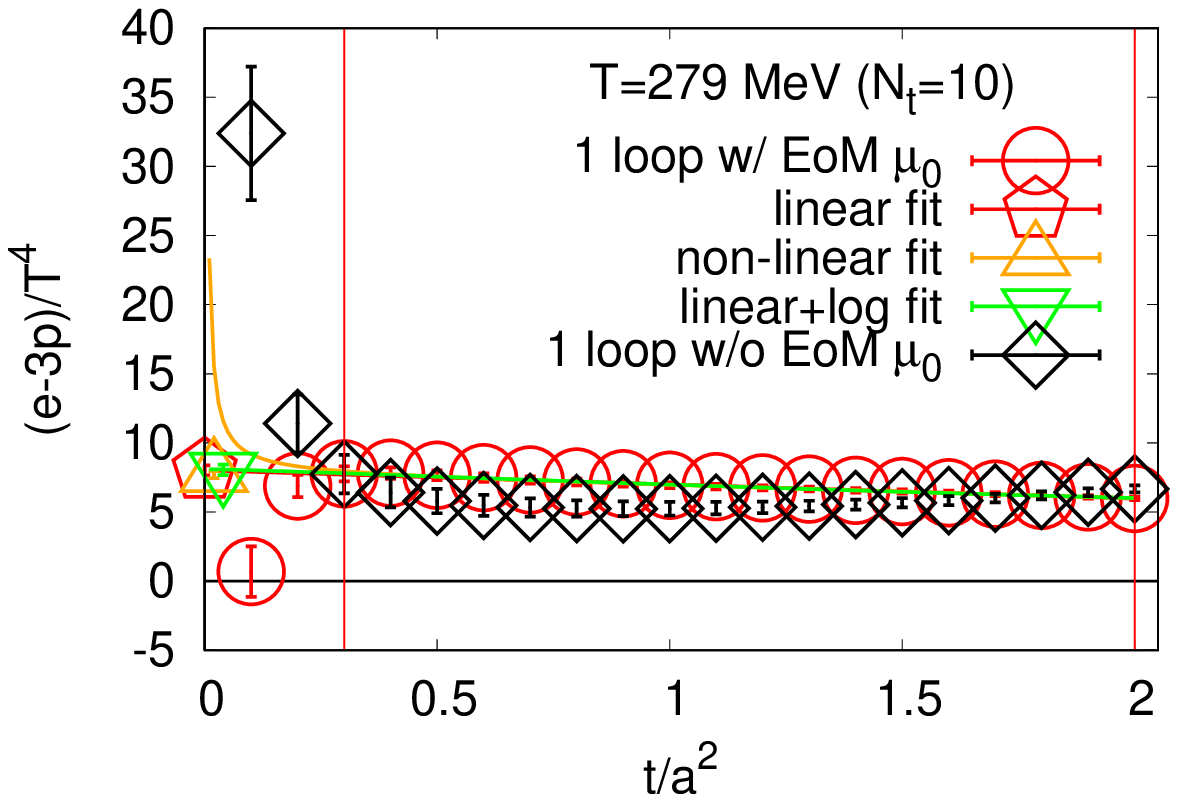}
  \includegraphics[width=4.9cm,trim=0 0 8mm 3mm,clip]{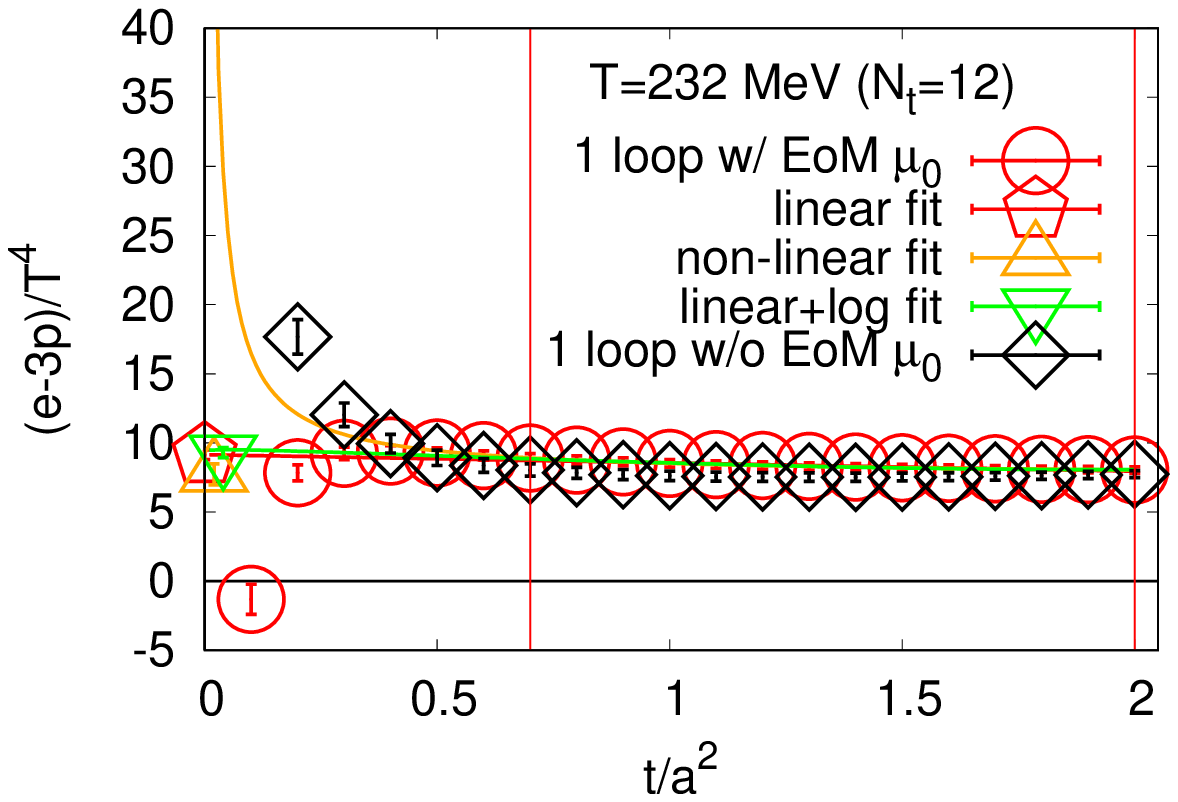}
  \includegraphics[width=5cm,trim=0 0 4mm 3mm]{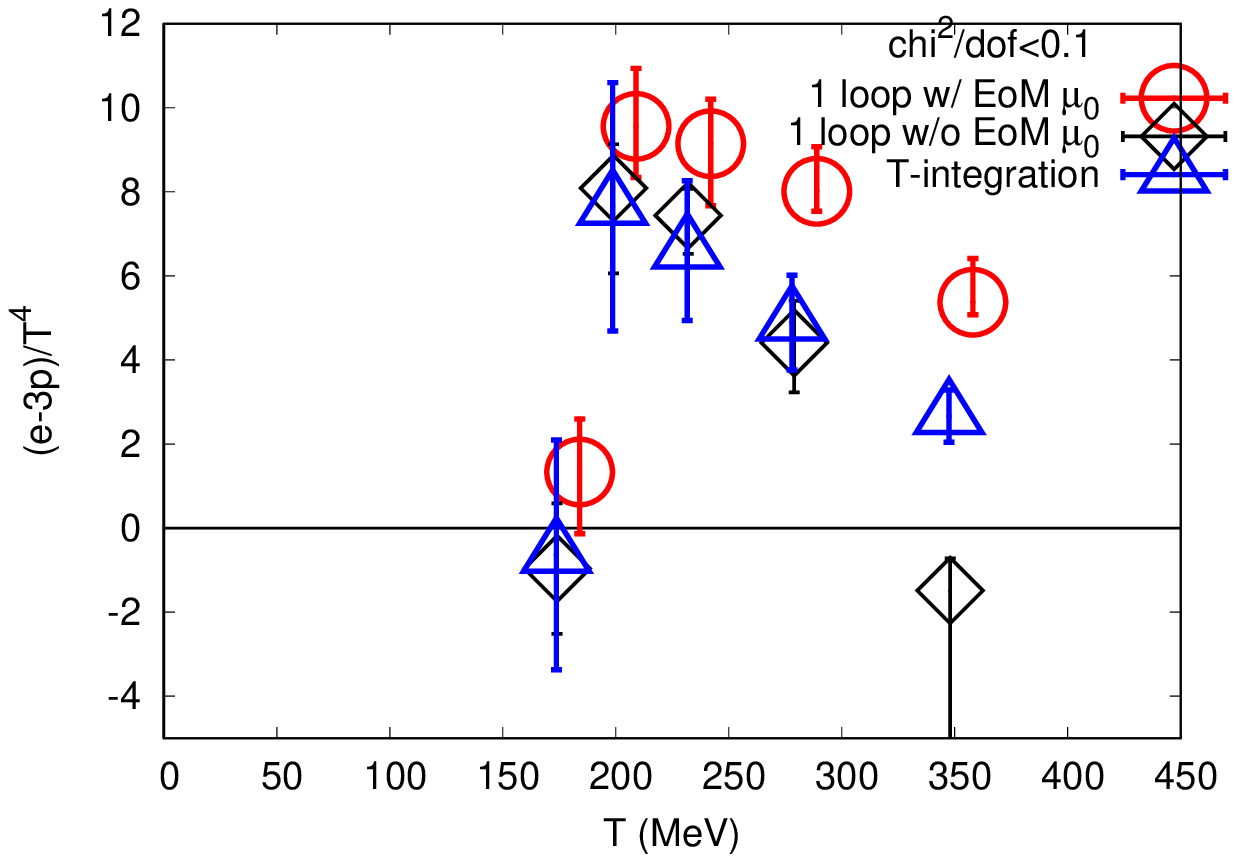}
  \vspace*{-3mm}
  \caption{
  Trace anomaly in QCD with slightly heavy $u$ and $d$ quarks using the $\mu_0$-scale.
  Comparison of results using the one-loop matching coefficients of~\cite{Makino:2014taa} without using the EoM and those using the one-loop  coefficients by Harlander \textit{et al.} using the EoM~\cite{HKL}. 
  \textbf{Left/Middle}: Trace anomaly at $T\simeq 232$ and 279 MeV as a function of the flow time. 
  \textbf{Right}: Trace anomaly in the $t\to0$ limit.
  Data at $T\simge350\,\mathrm{MeV}$ are contaminated with $O\!\left((aT)^2=1/N_t^2\right)$ lattice artifacts at~$N_t \simle 8$.
}
\label{fig:heavy2}
\end{figure}

In order to identify the effects of EoM more clearly, we compare, in Fig.~\ref{fig:heavy2}, the results of the trace anomaly with the one-loop matching coefficients of \cite{Makino:2014taa} without using the EoM and those with the one-loop part of \cite{HKL} in which the EoM is used. 
In the left and middle panels of Fig.~\ref{fig:heavy2}, we show the trace anomaly at $T\simeq 232$ and 279~MeV, respectively, as a function of the flow time.
The results for the trace anomaly in the $t\to0$ limit are shown in the right panel.
Comparing this with the right panel of Fig.~\ref{fig:heavy}, we confirm that the discrepancy at high temperatures are mainly due to the use of the EoM. 
We recall here that the EMT at $T\simge350$ MeV are contaminated with $O\!\left((aT)^2=1/N_t^2\right)$ lattice artifacts at $N_t \simle 8$~\cite{Taniguchi:2016ofw}. 
However, even with disregarding the data at $T\simge350$ MeV, we see some discrepancy at $T\simeq279$~MeV ($N_t=10$).
The $O\!\left((aT)^2\right)$ lattice artifacts will contaminate the EoM too.
Our results suggest that the EoM suffers from visible $O\!\left((aT)^2=1/N_t^2\right)$ discretization errors at $N_t \simle 10$.

\section{Summary and outlook}
\label{sec:summary}

We are studying thermodynamic properties of $2+1$ flavor QCD with improved Wilson quarks applying the SF\textit{t}X method based on the gradient flow. 
In our first study with $u$ and $d$ quarks slightly heavier than the real world, we found that the method works quite well in calculating various physical observables including the EMT, chiral condensate, chiral susceptibility, and topological susceptibility, even with the explicit violation of the chiral symmetry with Wilson quarks.
We are extending the study to QCD just at the physical point on a slightly less fine lattice. 
We found that the method works well also at the physical point when we adjust the renormalization scale appropriately.
Our preliminary results using the $\mu_0$-scale suggest that temperatures in $T \simeq 122$--146 MeV are in the critical region. 
To draw a definite conclusion, however, we need more statistics and data points at these temperatures.
We are currently generating configurations there.

In the second part of this report, we presented the results of our test of two-loop matching coefficients for EMT, which are recently calculated by Harlander \textit{et al.}~\cite{HKL}. 
For this test, we revisited the case of QCD with slightly heavy $u$ and $d$ quarks. 
A point to be noted here is that an EoM in the continuum limit is used by Harlander \textit{et al.}\ with which we reduce the number of independent quark operators in the EMT.
For the entropy density in which the use of the EoM has no effects,
we found that the results using the two-loop coefficients are well consistent with our previous results using one-loop coefficients. 
On the other hand, for the trace anomaly in which the EoM does affect, we found discrepancies between the one- and two-loop results at high temperatures (small $N_t$'s).
The main origin of the discrepancies was identified as the use of the EoM, by a direct comparison of the results of one-loop coefficients with and without using the EoM.
Our results suggest that the EoM suffers from large $O\!\left((aT)^2\right)=O\!\left(1/N_t^2\right)$ discretization errors at $N_t \simle 10$. 

Besides further studies just at the physical point, we are attempting to extend applications of the SF\textit{t}X method in various directions: 
shear and bulk viscosities in $2+1$ flavor QCD from two-point correlation functions of the EMT~\cite{Taniguchi:2017ibr},
end-point of first-order deconfining transition region in QCD near the quenched limit \cite{Shirogane}, 
PCAC quark masses~\cite{ABaba}, etc.
The $\mu_0$-scale as well as higher order coefficients may help improving these calculations too.

\vspace{2mm}
We are grateful to Prof.~R.V.~Harlander for valuable discussions. 
This work was in part supported by JSPS KAKENHI Grant Numbers 
JP19K03819, JP19H05146, JP18K03607, JP17K05442 and JP16H03982.
This research used computational resources of COMA, Oakforest-PACS, and Cygnus provided by the Interdisciplinary Computational Science Program of Center for Computational Sciences, University of Tsukuba,
K and other computers of JHPCN through the HPCI System Research Projects (Project ID:hp17208, hp190028, hp190036) and JHPCN projects (jh190003, jh190063), OCTOPUS at Cybermedia Center, Osaka University, and ITO at R.I.I.T., Kyushu University.
The simulations were in part based on the lattice QCD code set Bridge++ \cite{bridge}.


\end{document}